\documentclass[aps,prb,twocolumn,superscriptaddress]{revtex4}

\usepackage{graphicx,color}

\input epsf.sty
\usepackage{graphicx}

\begin{document}

\preprint{\today}


\title{
Role of magnetic chirality in polarization flip upon commensurate-incommensurate
magnetic phase transition in YMn$_{2}$O$_{5}$
}

\author{Shuichi Wakimoto
}
\affiliation{ Quantum Beam Science Directorate, Japan Atomic Energy Agency,
   Tokai, Ibaraki 319-1195, Japan }

\author{Hiroyuki Kimura}
\affiliation{ Institute of Multidisciplinary Research for Advanced Materials, 
   Tohoku University, Sendai 980-8577, Japan }

\author{Yuma Sakamoto}
\affiliation{ Institute of Multidisciplinary Research for Advanced Materials, 
   Tohoku University, Sendai 980-8577, Japan }

\author{Mamoru Fukunaga}
\affiliation{ Institute of Multidisciplinary Research for Advanced Materials, 
   Tohoku University, Sendai 980-8577, Japan }

\author{Yukio Noda}
\affiliation{ Institute of Multidisciplinary Research for Advanced Materials, 
   Tohoku University, Sendai 980-8577, Japan }

\author{Masayasu Takeda}
\affiliation{ Quantum Beam Science Directorate, Japan Atomic Energy Agency,
   Tokai, Ibaraki 319-1195, Japan }

\author{Kazuhisa Kakurai}
\affiliation{ Quantum Beam Science Directorate, Japan Atomic Energy Agency,
   Tokai, Ibaraki 319-1195, Japan }

\date{\today}

\begin{abstract}

We have performed simultaneous measurements of magnetic chirality by using 
polarized neutrons and electric polarization along the $b$-axis of 
single crystals of YMn$^{4+}$(Mn$_{1-x}$Ga$_{x}$)$^{3+}$O$_{5}$ with $x=0.047$ 
and $0.12$, in which nonmagnetic Ga-ions dilute Mn$^{3+}$ spins.
The $x=0.047$ sample  exhibits high-temperature incommensurate (HT-ICM), 
commensurate (CM), and  low-temperature incommensurate (LT-ICM) 
magnetic phases in order of decreasing temperature, whereas the $x=0.12$ sample 
exhibits only HT-ICM and LT-ICM phases.  Here, the CM and LT-ICM phases are 
ferroelectric and weak-ferroelectric, respectively.
Measurements conducted under zero field heating after various field-cooling 
conditions evidence that 
the microscopic mechanisms of the spin-driven 
ferroelectricity in the CM and LT-ICM phases are different:
the magnetic chirality of Mn$^{4+}$ cycloidal spins 
plays a dominant role in the LT-ICM phase, whereas the 
magnetic exchange striction by the Mn$^{4+}$-Mn$^{3+}$ chain plays a 
dominant role in the CM phase.
The polarization of YMn$_{2}$O$_{5}$ flips upon CM to LT-ICM phase transition 
because the ferroelectricity driven by the magnetic chirality and the exchange 
striction provides opposite directions of polarization.

\end{abstract}


\maketitle

Many of the multiferroic compounds that have attracted much attention in recent years 
exhibit a giant magnetoelectric effect owing to the strong coupling between 
magnetism and ferroelectricity.  The presence of this effect makes these 
multiferroic compounds 
potential candidates for industrial device applications.  In addition, the 
coupling mechanism underlying the giant magnetoelectric effect is one of the 
central issues in this research field. 

Multiferroic material $R$Mn$_{2}$O$_{5}$ (R: rare earth, Bi, and Y)~\cite{Hur2004} 
provides 
a unique opportunity for studying the coupling mechanism between magnetism 
and ferroelectricity.  
$R$Mn$_{2}$O$_{5}$,
which has an orthorhombic structure with the $Pbam$ space group,
consists of edge-shared Mn$^{4+}$O$_{6}$ octahedral chain along the $c$-axis 
and pairs of Mn$^{3+}$O$_{5}$ pyramids connecting the Mn$^{4+}$O$_{6}$ octahedral 
chains.~\cite{Alonso1997}
Frustrated spin exchange interactions result 
in noncollinear magnetic structures and complex magnetic phases.~\cite{Chapon2004}
$R$Mn$_{2}$O$_{5}$ generally exhibits a magnetic phase transition from a paramagnetic 
to an incommensurate magnetic (HT-ICM) phase  at $T_{N1}$, which is typically 
$\sim 45$~K, followed by a successive magnetic phase transition into a commensurate 
(CM) phase at $T_{CM}$, typically a few kelvins below $T_{N1}$.  Then, at lower
temperature, the magnetic structure becomes incommensurate again (a phase 
referred to as LT-ICM) at $T_{N2}$.~\cite{Kagomiya2001,Noda2008}  
These three magnetic phases exhibit different 
dielectric properties.  The HT-ICM phase is paraelectric, whereas the CM phase 
exhibits spontaneous polarization concomitantly, and therefore, the CM phase is 
ferroelectric (FE).  The LT-ICM phase is still ferroelectric, however the 
polarization is suppressed below the CM to LT-ICM phase transition temperature.  
Thus, 
the LT-ICM phase is recognized as weak ferroelectric (WFE).
Remarkably, a rich variety of magnetoelectric effects have been identified in 
the LT-ICM phase of $R$Mn$_{2}$O$_{5}$.  Such a variety of magnetic and dielectric 
phases implies a complicated coupling mechanism in this system.~\cite{Kim2008}

For the specific case of YMn$_{2}$O$_{5}$, $T_{N1}=44$~K, $T_{CM}=38$~K, and 
$T_{N2}=20~K$.
The magnetic modulation vectors for the HT-ICM, CM, and LT-ICM phases are 
$k = (0.254, 0, 0.484)$, $(0.25, 0, 0.5)$, and $(0.292, 0, 0.48)$, respectively.
The magnetic structure in the CM phase determined precisely by single crystal 
neutron diffraction is reported in Ref.~\onlinecite{Kimura2007}.  
Characteristic is a spiral modulation of spins along the $c$-axis. 
The magnetic structure in the LT-ICM phase can be considered as a similar spiral 
with incommensurate modulation period from the polarized neutron diffraction.~\cite{Kim2008}
Spontaneous polarization along the $b$-axis appears concomitantly with the CM 
phase.  Remarkably it is known that the polarization flips upon the CM to 
LT-ICM phase transition when the sample is poled at the CM 
phase.~\cite{Inomata1996,Kagomiya2003,Chaudhury2009} 

The exchange striction involving both Mn$^{3+}$ and Mn$^{4+}$ spins has been proposed
for explaining magnetically driven ferroelectric polarization in $R$Mn$_{2}$O$_{5}$, 
based on powder magnetic structural analyses by using neutron 
diffraction.~\cite{Chapon2006,Blake2005}  In this model,
the polarization is proportional to the scalar product of neighboring spins, 
${\rm\bf P} \propto {\rm\bf S}_i \cdot {\rm\bf S}_j$. 
This model can reproduce the magnetic polarization in both CM and LT-ICM phases 
in YMn$_{2}$O$_{5}$.~\cite{Chapon2006} 

Noncollinear spin arrangements or cycloidal spin structures 
can break the inversion symmetry of the lattice through Dzyaloshinskii-Moriya 
antisymmetric interactions, which results in spontaneous 
polarization.~\cite{Katsura2005,Sergienko2006}  
In this case, the polarization is proportional to the vector product of 
neighboring spins, ${\rm\bf P} \propto {\rm\bf S}_i \times {\rm\bf S}_j$.   
This relation has been confirmed in many multiferroic compounds 
such as orthorhombic $R$MnO$_{3}$, 
MnWO$_{4}$, 
$R$Mn$_{2}$O$_{5}$ and Ni$_{2}$V$_{2}$O$_{8}$ 
by using polarized neutron 
diffraction.~\cite{Yamasaki2007,Sagayama2008,Fukunaga2009,Cabrera2009}
Magnetic structural analyses using single-crystal neutron diffraction for 
various $R$Mn$_{2}$O$_{5}$ compounds confirmed the cycloidal spin structure of Mn$^{4+}$ 
spins propagating along the $c$-axis.~\cite{Noda2006,Kimura2007,Vecchini2008,Kim2008,Radaelli2009}  
This structure is also consistent 
with $b$-axis polarization.

\begin{figure}
\includegraphics[width=8cm]{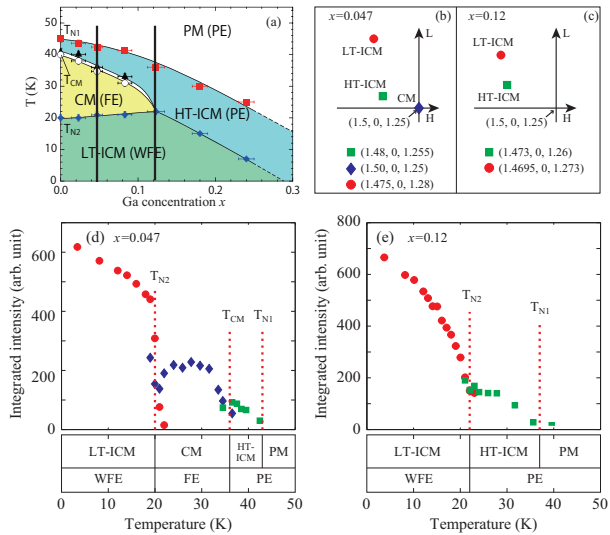}
\caption{(Color online) (a) Magnetic and dielectric phase diagram for 
YMn$^{4+}$(Mn$_{1-x}$Ga$_{x}$)$^{3+}$O$_{5}$ adopted from Ref.~\onlinecite{Kimura2013}.
Vertical lines represent $x=0.047$ and $0.12$, which are used for the present 
study.
(b) and (c) show CM and ICM peak geometry in the (1.5, 0, 1.25) 
zone for $x=0.047$ and $0.12$, respectively.
(d) and (e) show temperature dependence of magnetic peak 
intensities for $x=0.047$ and $0.12$, respectively, measured using unpolarized neutrons 
at the position specified in (b) and (c).  Squares, diamonds, and circles represent 
data for HT-ICM, CM, and LT-ICM peaks, respectively.
}
\end{figure}

Very recently, a magnetic and dielectric phase diagram was reported for 
YMn$^{4+}$(Mn$_{1-x}$Ga$_{x}$)$^{3+}$O$_{5}$ [Fig. 1(a)] wherein Mn$^{3+}$ ions are 
selectively substituted by nonmagnetic Ga$^{3+}$ ions.~\cite{Kimura2013}  The Ga$^{3+}$ 
substitution does not affect the Mn$^{4+}$-Mn$^{4+}$ exchange interaction, and hence nor the 
cycloidal structure of Mn$^{4+}$ spins, but it does affect the Mn$^{4+}$-Mn$^{3+}$ exchange 
interaction.  
Therefore, the Ga$^{3+}$ substitution should dilute the contribution of the exchange striction 
mechanism to the spontaneous polarization.
The phase diagram indicates that as Ga concentration increases, the CM (FE) phase 
is diminished below $x=0.12$.  The LT-ICM (WFE) phase survives up to higher concentrations.
This phase diagram suggests that ferroelectricity in the CM phase 
originates from the exchange striction, whereas weak ferroelectricity in the 
LT-ICM phase originates from the cycroidal structure of Mn$^{4+}$ spins.  
A similar scenario has been hypothesized in Ref.~\onlinecite{Fukunaga2010} from 
dielectric measurements.
So far, however, direct experimental evidence for this scenario is missing.
In this Letter, we report results of simultaneous measurements of magnetic chirality 
and dielectric polarization in various cooling conditions and provide direct 
experimental evidence that the magnetic chirality of Mn$^{4+}$ cycloidal spins 
plays a dominant role in the LT-ICM phase, whereas the 
magnetic exchange striction by the Mn$^{4+}$-Mn$^{3+}$ chain plays a 
dominant role in the CM phase.

We carried out simultaneous measurements 
of magnetic chirality by using polarized neutron diffraction and electric polarization 
for Ga-substituted YMn$_{2}$O$_{5}$ and studied the relation between chirality 
and polarization in each magnetic phase. 
We chose Ga concentrations of $x=0.047$ and $0.12$.  These concentrations are 
specified by vertical bars in Fig. 1(a).  Single crystals are grown in 
the same manner as those reported in Ref.~\onlinecite{Kimura2013}. 
Magnetic phase transitions in these samples are verified using neutron diffraction, 
performed at the triple-axis spectrometer TAS-1 installed at the JRR-3 reactor, Tokai, 
Japan. 
Figures 1(d) and 1(e) show temperature dependence of magnetic peak intensity 
for both samples.  The measured ${\rm\bf Q}$ positions are schematically shown 
in Figs. 1(b) and 1(c).  The detailed change of the magnetic propagation vector has been 
reported in Ref.~\onlinecite{Kimura2013}. 
Consistently with the phase diagram in Fig. 1(a), the $x=0.047$ sample 
exhibits HT-ICM, CM, and LT-ICM phases during cooling, whereas the $x=0.12$ sample 
has a single magnetic phase transition from HT-ICM to LT-ICM.
Therefore, the former sample has both FE and WFE phases, and the latter sample 
has only the WFE phase.

\begin{figure}
\includegraphics[width=8cm]{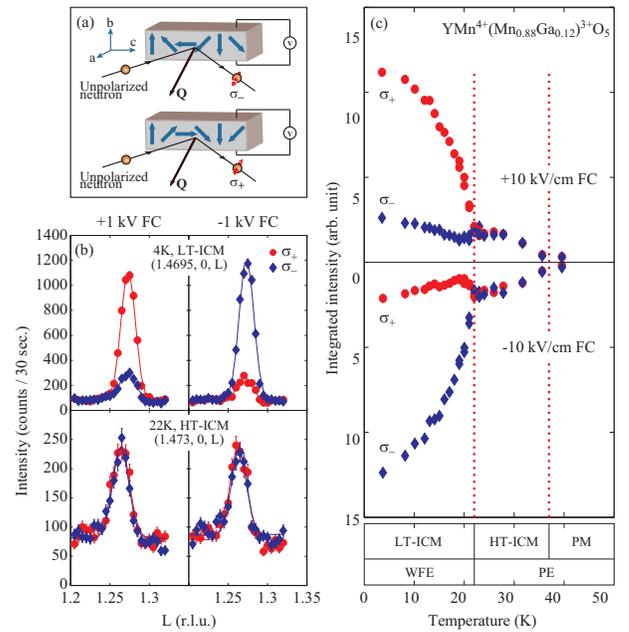}
\caption{(Color online) (a) Schematic drawings of the experimental geometry 
of polarized neutron diffraction.  (b) $\sigma_{+}$ and $\sigma_{-}$ 
profiles of LT-ICM peak (top) and HT-ICM (bottom) measured after cooling 
with electric fields of $+10$~kV/cm (left) and $-10$~kV/cm (right) for 
$x=0.12$.  (c) Temperature dependence of $\sigma_{+}$ (circles) and 
$\sigma_{-}$ (diamonds).
}
\end{figure}

Simultaneous measurements of polarized neutron diffraction and electric 
polarization were carried out also at TAS-1.  
The spin polarizations of the incident and diffracted neutrons, ${\bf P_i}$ and 
${\bf P_f}$, are connected by the following formula when involving only magnetic scattering 
cross section from a magnetic structure:~\cite{Blume1963}
${\bf P_f} \sigma = -{\bf P_i}({\bf M}^{*} \cdot {\bf M}) + {\bf M}^{*}({\bf P_i} \cdot {\bf M}) + ({\bf P_i} \cdot {\bf M}^{*}){\bf M} + i({\bf M}\times{\bf M}^{*})$,
where ${\rm\bf M}$ is a magnetic structure factor
${\bf M} \propto \sum_{j}{{\bf S}_{j\perp}e^{i{\bf Q}\cdot{\bf r}_j}}$,
and {\bf M}$^{*}$ is its complex conjugate.
Here, $\sigma$, and ${\bf S}_{j\perp}$ are the total neutron 
cross section and a component of the $j-$th 
spin perpendicular to the scattering vector {\bf Q}.  
The fourth term expresses the chiral structure contribution, 
which gives the final polarization parallel to {\bf Q}.
We kept the incident neutrons unpolarized and analyzed the polarization of 
the diffracted neutrons using a Heusler analyzer with a spin flipper in front 
of the analyzer.  A guide field around the sample was kept parallel to 
{\bf Q} by a Helmholtz coil.  
Thus we analyze the final polarization ${\bf P_f}$ parallel to {\bf Q}.
Neutron cross sections with spin (+), $\sigma_{+}$, and (-), $\sigma_{-}$, 
can be measured with the spin flipper off (non-spin-flip 
channel) and the spin flipper on (spin-flip channel), respectively. 
From the guide field 
configuration, the polarizations of $\sigma_{+}$ and $\sigma_{-}$ are antiparallel 
and parallel to {\bf Q}, respectively.  This is schematically shown in Fig. 2(a).
Now {\bf Q}-parallel component of ${\bf P_f}$ is $(\sigma_{+}-\sigma_{-})/(\sigma_{+}+\sigma_{-})$, 
while total neutron cross section can be written as $\sigma = \sigma_{+}+\sigma_{-}$.  
In the present experimental setup, the previous formula can be simplified by extracting 
a {\bf Q}-parallel component as: 
$\sigma_{+}-\sigma_{-} = i({\bf M}\times{\bf M}^{*})_{\rm\bf Q}$.  
Therefore, the difference in cross sections, $\sigma_{+}-\sigma_{-}$, records 
the magnetic chirality $i({\bf M}\times{\bf M}^{*}$) directly.

\begin{figure}
\includegraphics[width=8cm]{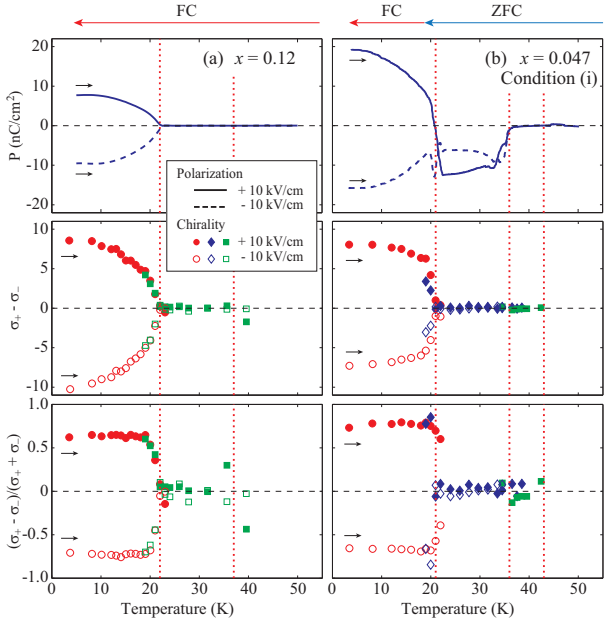}
\caption{(Color online) (a) Temperature dependence of electric polarization 
$P$, $\sigma_{+}-\sigma_{-}$ as a measure of magnetic chirality, and 
normalized chirality $(\sigma_{+}-\sigma_{-})/(\sigma_{+}+\sigma_{-})$ for 
(a) $x=0.12$ measured during the ZFH process after the FC process and (b) $x=0.047$ 
measured after ZFC down to 18~K followed by FC down to 4~K.}
\end{figure}

First, let us discuss the simpler case shown in Fig. 1(e), that is, the phase 
transition from HT-ICM to LT-ICM for $x=0.12$.  
Figure 2 summarizes a typical data set for the $x=0.12$ sample.  The crystal was 
placed with $a$- and $c$-axes horizontal.  Electrodes were attached so that an 
electric field can be applied along the $b$-axis.  
Figure 2(b) shows cross sections $\sigma_{+}$ and $\sigma_{-}$ measured at the 
{\bf Q} positions specified in Fig. 1(c).  Measurements were taken during a heating 
process without an electric field (zero-field heating, ZFH)
after a cooling process with an applied electric field (field cooling, 
FC) of 10~kV/cm from 55 to 4~K.  
At 22~K in the HT-ICM phase, which is paraelectric, there is no difference 
between $\sigma_{+}$ and $\sigma_{-}$.  This means that there is no magnetic 
chirality (or imbalance of chiral domains) in the paraelectric field.   
In contrast, at 4~K in the LT-ICM WFE phase, there is a difference between 
$\sigma_{+}$ and $\sigma_{-}$.  The magnitudes of these cross sections are 
reversed by changing the sign of the electric field.  Therefore, the magnetic 
chirality flips when the electric polarization flips.  
We analyzed the peak profiles by fitting to a Gaussian function and derived 
the integrated intensity.  
The detailed temperature dependence of the integrated intensities of $\sigma_{+}$ and 
$\sigma_{-}$ is shown in Fig. 2(c).  The difference only appears in the 
LT-ICM WFE phase.

\begin{figure}
\includegraphics[width=8cm]{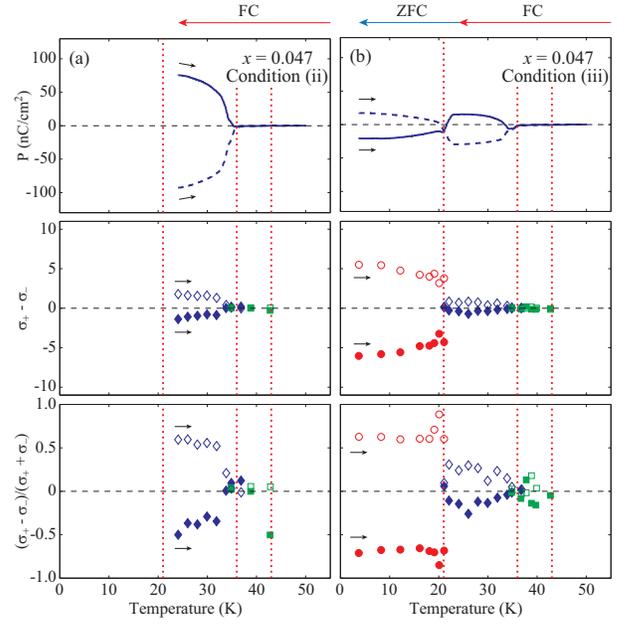}
\caption{(Color online) Analogous data plots to those of Fig. 3 for $x=0.047$ measured 
after (a) FC down to 24~K, and (b) FC down to 24~K followed by ZFC down to 4~K.}
\end{figure}

The magnetic chirality, characterized as the difference $\sigma_{+}-\sigma_{-}$, is 
shown in Fig. 3(a), together with the electric polarization $P$ measured using 
pyrocurrent 
observed in a ZFH process.  
It is clearly shown that these two parameters follow identical 
temperature dependence in the LT-ICM phase.  This behavior is exactly the same 
as that observed in TbMnO$_{3}$.~\cite{Yamasaki2007}  
In Fig. 3(a), we show the normalized chirality 
$(\sigma_{+}-\sigma_{-})/(\sigma_{+}+\sigma_{-})$ in the bottom panel.  
This parameter remains nearly constant up to $T_{N2}$ at $\sim 0.6$.  
Note that the normalized chirality should be less than 1 even when the 
magnetic structure 
has a single chiral domain in case of YMn$_2$O$_5$ since Mn spins 
form a cycloidal structure with elongated magnetic moment along the 
$a$-axis,~\cite{Kimura2007} 
and thence ${\bf S}_{j\perp}$ forms an ellipsoid at {\bf Q} we measure.
(This fact corresponds to appreciable $\sigma_{-}$ (or $\sigma_{+}$) of the 
$+10$~kV/cm (or $-10$~kV/cm) data at 4~K in Fig. 2(b).)
We calculated the normalized chirality at $(1.5, 0, 1.25)$ using the magnetic 
structure of the CM phase in YMn$_{2}$O$_{5}$ reported in 
Ref.~\onlinecite{Kimura2007} with single chirality.  Calculated value is $0.52$ 
which is reasonably consistent with the observed value $0.60$.
This means that, through the FC process, 
the sample was fully poled and contained a single magnetic domain.
It should be noted here that, in the LT-ICM phase of this sample, polarization 
and chirality have the same sign.

As a second case, we made similar measurements for the $x=0.047$ sample. 
Because this sample has 
two dielectric phases, FE and WFE, we employed three different FC conditions: 
(i) zero-field cooling (ZFC) from 55 to 18~K, followed by FC from 18 to 4~K, 
(ii) FC from 55 to 24~K, and
(iii) FC from 55 to 24~K, followed by ZFC from 24 to 4K.
Figure 3(b) shows the plots of polarization $P$, chirality $(\sigma_{+}-\sigma_{-})$, 
and normalized chirality $(\sigma_{+}-\sigma_{-})/(\sigma_{+}+\sigma_{-})$ for 
condition (i).
This condition corresponds to poling of the LT-ICM WFE phase only. 
In this case, $P$ and the chirality in the LT-ICM phase have the same sign, and 
they are reversible by changing the electric field.  Thus, the behavior in the LT-ICM 
phase is consistent with that in the $x=0.12$ sample.
In the ZFH process, there remains a finite polarization in the CM phase, 
likely resulting from an incipient imbalance of ferroelectric domains; 
nevertheless, the chirality disappears.  This shows that the ferroelectricity 
in the CM phase does not require the magnetic chirality.

In contrast, when the sample is poled in the CM-FE phase [condition (ii), Fig. 4(a)], 
both $P$ and the chirality appear in the CM phase.  Here $P$ in the CM phase 
is about four times larger than that in the LT-ICM phase in Fig. 3(b).  (Note the difference 
in the vertical scales between Figs. 3 and 4.)
$P$ and the chirality are again reversible by changing the electric field as a normal character 
of ferroelectricity; however, they have opposite signs: for example, $P$ is 
positive with +10~kV/cm but the chirality is negative.

In both conditions (i) and (ii) [Fig. 3(b) and 4(a)], 
the normalized chirality reaches $\sim 0.6$; that is, both poled 
CM and LT-ICM phases have a single magnetic domain but opposite chirality. 
This means that nearly the same magnetic structure gives opposite spontaneous 
polarization in the CM and LT-ICM phases. 
This is clearly shown in Fig. 4(b), which shows the data in condition (iii).  
The sample is once poled in the CM phase, then zero-field-cooled to 4K. 
In this case, the magnetic structure poled in the CM phase is preserved 
down to 4~K, and therefore, the system gives nearly identical values of  $P$ and 
normalized chirality in the opposite direction to that of the condition (i) 
[Fig. 3(b)] in the LT-ICM phase.
Thus, the chiral domain poled in the CM phase is preserved in the LT-ICM phase 
with the ZFC process.
In contrast, in condition (i) in Fig.3(b), the chirality vanishes 
at $T_{N2}$.  This indicates that the chiral domain poled in 
the LT-ICM phase is not preserved in the CM phase even with the ZFH process.  
This is also consistent with the observation in Fig. 4(b) showing that the chirality 
vanishes at $T_{N2}$ in the ZFH process.
These facts indicate that the ferroelectricity in the CM and LT-ICM phases has 
different origins although it arises from nearly identical magnetic structures.  
Moreover, the results in Fig. 3 strongly support that  magnetic chirality 
is the origin of weak ferroelectricity in the LT-ICM phase.

\begin{figure}
\includegraphics[width=6cm]{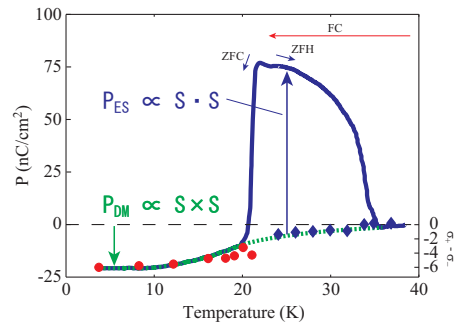}
\caption{(Color online)
Total polarization (solid line) together with the magnetic 
chirality in YMn$_{2}$O$_{5}$ when the system is poled at the CM phase. 
Diamonds and circles are plotted $\sigma_{+}-\sigma_{-}$ of the CM phase in 
Fig. 4(a) and those of LT-ICM in Fig. 4(b), respectively.
$P_{DM}$ contribution is shown by a dotted line.
}
\end{figure}

Based on the above, the total electric polarization in YMn$_{2}$O$_{5}$ can be 
expressed as a summation of two components: $P = P_{ES} + P_{DM}$, where 
$P_{ES}$ ($\propto {\rm \bf S_i} \cdot {\rm \bf S_j}$) is dominant in the CM 
phase and $P_{DM}$ ($\propto {\rm \bf S_i} \times {\rm \bf S_j}$) is dominant 
in the LT-ICM phase.
This is schematically shown in Fig. 5. 
When the system is poled in the CM phase, it exhibits a single magnetic domain 
and $P_{ES}$ has positive polarization.  However, this magnetic structure 
possesses a chirality that gives negative $P_{DM}$.  
Upon ZFC across the CM to the LT-ICM phase transition, 
${\rm \bf S_i} \cdot {\rm \bf S_j}$ information is lost by the incommensurate 
modulation period
while ${\rm \bf S_i} \times {\rm \bf S_j}$ information is preserved.
Therefore, only $P_{DM}$ survives and the electric 
polarization flips.
This model is also evidenced by that $P$ in the CM phase in Fig. 4(b) is much 
smaller than Fig. 4(a).
In case of Fig. 4(b), the sample once enters the LT-ICM phase after poled at the 
CM phase, and bulk ferroelectricity by ${\rm \bf S_i} \cdot {\rm \bf S_j}$ is 
diminished by the incommensurate structure
of LT-ICM.  Thus, the ferroelectricity 
does not fully recover when the system reenter the CM phase on ZFH.

In summary, we performed simultaneous measurements of magnetic chirality 
and electric polarization for YMn$^{4+}$(Mn$_{1-x}$Ga$_{x}$)$^{3+}$O$_{5}$ in 
which nonmagnetic Ga$^{3+}$ dilutes the contribution of 
${\rm \bf S_i} \cdot {\rm \bf S_j}$ to the ferroelectricity.
Polarization, magnetic chirality, and normalized magnetic chirality measured 
under various field-cooling conditions evidence that 
the ferroelectric polarizations in the CM and LT-ICM phases are dominated by 
components $P_{ES}$ ($\propto {\rm \bf S_i} \cdot {\rm \bf S_j}$) and $P_{DM}$ 
($\propto {\rm \bf S_i} \times {\rm \bf S_j}$), respectively.
Owing to the similar magnetic and dielectric properties of $R$Mn$_{2}$O$_{5}$, 
this phenomenon is likely to be valid for the most family compounds 
of this system.
The signs of $P_{ES}$ and $P_{DM}$ may depend on compounds with different $R$-ions.
In the specific case of $R$=Y, 
the polarization flip upon the CM to LT-ICM phase transition 
can be accounted for by the different signs of $P_{ES}$ and $P_{DM}$.

This work is supported by Grant-In-Aid for Scientific Research (B) No. 16340096, 
and Scientific Research (A) No. 21244051 and Grant-In-Aid for Scientific Research on 
Priority Areas ``Novel States of Matter Induced by Frustration" No. 19052001.




\end{document}